\documentclass{ifae}
\usepackage{epsfig,subfigure}

\makeatletter
\def\spig{\kern.15em\raise1.2ex\hbox{$|$}\kern-.48em\to} 
\def\anti#1{\mathpalette{\@anti}{#1}#1}
\def\@anti#1#2{\sbox0{$#1#2$}
  \makebox[0pt][l]
    {$#1\kern.30\ht0\overline{\kern-.35\ht0\phantom{#2}\kern-.1ex}$}}
\makeatother
\def\hlinestrut{\raise.3ex\hbox to 0pt{\strut}}

\begin{document}
\begin{header}
  \title{Status and prospects of charm physics: a few considerations}

  \begin{Authlist}
    S.~Malvezzi\Iref{mi}

   \Affiliation{mi}{INFN, Milano, Italy}
  \end{Authlist}

\begin{abstract}
The goals of quark flavor physics are to test the consistency of the Standard
Model (SM) description of quark mixing and CP violation, to search for evidence
of New Physics, and to select between New Physics scenarios that might be
initially uncovered at the LHC. This will require a range of measurements at
the percent level or better in the flavor-changing sector of the SM. Charm
represents a unique candidate for these studies. Progress in charm physics has
been prodigious over the last twenty years; it comes both from fixed-target and
collider ($e^+e^-$ and $p\bar p$) experiments, and can guide us toward future
investigations. To fully exploit the potentiality of flavor physics in
indicating New Physics, non-perturbative strong-interaction dynamics has to be
dealt with. Complications in the determination of CP phases due to strong
final-state interaction (FSI) have been uncovered; studies of charm-meson
decays, mainly through Dalitz-plot analyses, have experimentally confirmed  the
relevance of FSI phases. Analogously, determination of Cabibbo-suppressed CKM
matrix elements and $D\bar D$ mixing parameters will require an understanding
of strong-interaction effects among the light hadrons produced in heavy-meson
decays. The same complications will affect the beauty sector and will need to
be kept under control to search for New Physics effects.
 The role of charm in the search for New Physics, the lessons
learnt so far and warnings for the future will be discussed.
\end{abstract}

\end{header}
%

\section{Beyond the Standard Model; the clue from charm}

The success of the Standard Model in describing the experimental information to
date suggests that possible deviations have to be pursued either at high-energy
scales or as small effects in low-energy variables. In the Standard Model, CP
violation (CPV) and flavor-changing neutral currents (FCNC) are expected to be
small in charm decay, while lepton-family number violation and lepton-number
violation are forbidden. Anomalously large rates for rare decays would imply
non-Standard Model tree-level FCNC diagrams or non-Standard Model contributions
to higher-order loop diagrams, which make them sensitive to high-mass gauge
bosons and fermions and allow for probing particle states and mass scales not
directly accessible \cite{Schwartz}. The study of FCNC has been mainly
dedicated to transitions such as $s\to d l^+l^-$, $s\to d\nu\bar\nu$, $b\to
s\gamma$ and $b\to s l^+ l^-$, as well as phenomena such as $K^0$--$\bar K^0$
and $B^0$--$\bar B^0$ mixing. The analogous FCNC processes in the charm sector
have been investigated less. The Standard-Model expectations for both
$D^0$--$\bar D^0$ mixing and FCNC are very small, while extensions of the
Standard Model may enhance the contributions, which can then be orders of
magnitude larger. In the Standard Model, the $D$ system is not as sensitive to
CP as the $K$ and $B$ mesons. In this case too, the small predicted effects
could leave open a window to the observation of New-Physics effects. Charm
decays may be the only window onto such New Physics since it is possible that
the mechanism responsible will only couple to up-type quarks. More
specifically, non-Standard-Model forces might exhibit very different patterns
for the up and down classes of quarks \cite{BigSan}. Charm decays are the only
up-type decays that afford a probe of such physics: non-strange light-flavor
hadrons do not allow for oscillations and top-flavored hadrons do not even form
in a practical way. In Table~\ref{tab:expnow} experiments that have already
published results on charm mixing, rare decays and CPV searches 
are listed, along with their characteristics and a reference sample.
\begin{table}[htbp]
  \centering
  \caption{A
  comparison of the LEP experiments, CDF,
  E791, FOCUS, CLEO, BaBar and Belle.
  $K^-\pi^+$ is the
   number of reconstructed $D^0\to K^-\pi^+$ used
   in the published measurements. $\sigma_t$ denotes the
   proper time resolution for charm hadrons.}
  \label{tab:expnow}
  \begin{tabular}{ccccccc}
   \hline \hlinestrut
   &  & Fixed Target & &  $e^+e^-$  & &  $p\bar p$ \\
   \hline \hlinestrut
   & E791 & FOCUS &LEP& CLEO& BaBar/Belle& CDF \\
   \hline \hlinestrut
   Beam & Hadron & Photon & $e^+e^-\to Z^0$ & $e^+e^-$ & $e^+e^-$ & $p\bar p$\\
   $K^-\pi^+$ & $\sim 2 \times 10^4$ & $\sim 2 \times 10^5$ & $\sim10^4/$expt.
   & $\sim 2 \times 10^5$ & $\sim 10^6$ & $\sim 5 \times 10^5$ \\
   $\sigma_t$ & $ \sim 40$\,fs &  $ \sim 40 $\,fs &  $ \sim 100 $\,fs &
   $ \sim 140 $ &  $ \sim 160$\,fs &  $ \sim 50$\,fs \\
   \hline
  \end{tabular}
\end{table}
In Table ~\ref{tab:kpi} the expected data sets for existing experiments and
proposed facilities are reported \cite{BurShi}.
\begin{table}[htbp]
  \centering
  \caption{Expected $D^0\to K^-\pi^+$ samples for the existing
    experiments and proposed facilities. $K^-\pi^+$ is the number of of reconstructed $D^0\to K^-\pi^+$ in the full
    data set.}
  \label{tab:kpi}
  \begin{tabular}{ccc}
    \hline \hlinestrut
    Experiment & Full data set & $K^-\pi^+$ reconstructed        \\
    \hline \hlinestrut
    BaBar       &  500  fb$^{-1}$    & $6.6 \times 10^6$         \\
    Belle       &  500  fb$^{-1}$    & $6.6 \times 10^6$         \\
    CDF         &  4.4  fb$^{-1}$    & $ 30 \times 10^6$         \\
    CLEO-c      &  3.0  fb$^{-1}$    & $5.5 \times 10^5$         \\
    BES III     &  30.0 fb$^{-1}$    & $5.5 \times 10^6$         \\
    BTeV        &                    & $\sim 6 \times 10^8/10^7$      \\
    SuperBaBar  &  10.0 ab$^{-1}$    & 1.3 $\times 10^8 /10^7$\,s \\
    SuperKEKB   &   2.0 ab$^{-1}$    & 2.5 $\times 10^7 /10^7$\,s \\
    \hline
  \end{tabular}
\end{table}

\subsection{Mixing}

\begin{figure}[hbtp]
  \centering
  \epsfig{file=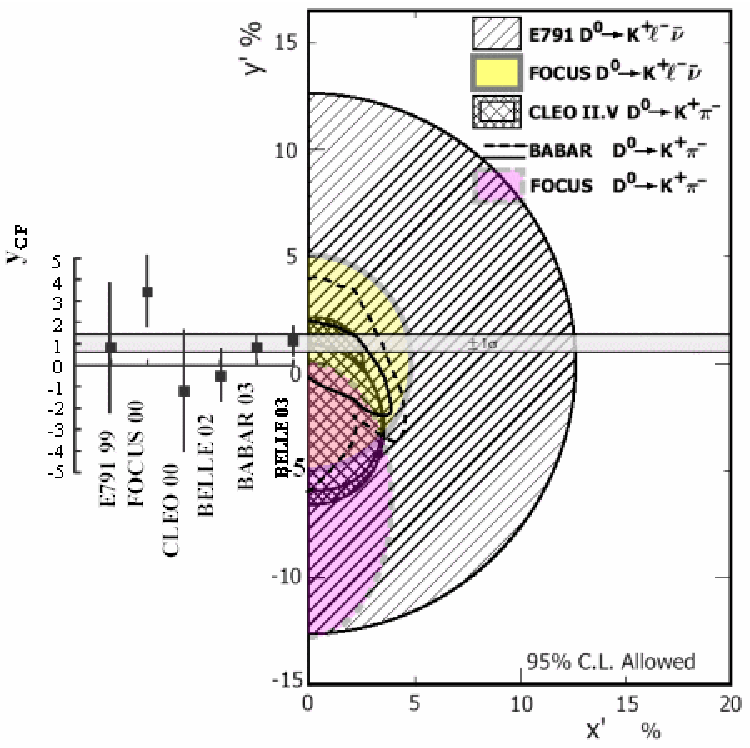,width=0.45\linewidth}
  \epsfig{file=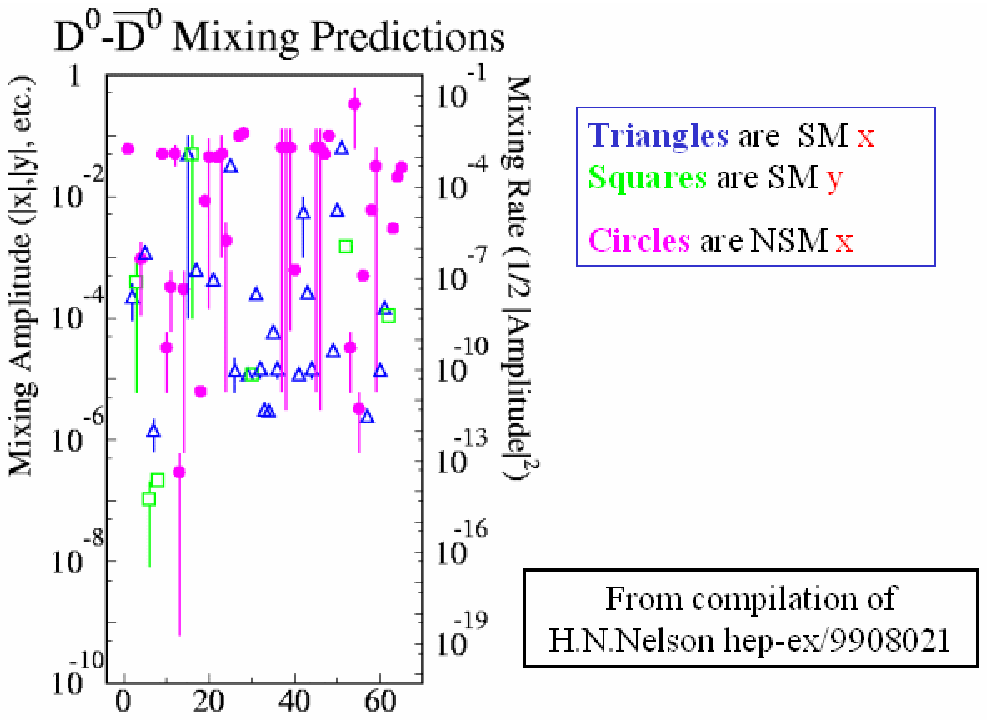,width=0.54\linewidth}
  \caption{Mixing scenario.}
  \label{fig:mix04}
\end{figure}

$D^0-\bar D^0$ mixing in the Standard Model occurs because the two eigenstates
$D^0$ and $\bar D^0$ are not mass eigenstates. The probability that a $D^0$
meson produced at $t=0$ decays as a $\bar D^0$ at time $t$ is given by
\begin{equation}
P(D^0\to\bar D^0) = \frac{1}{4} e^{-\Gamma_1t}\left\{ 1-2 e^{\frac{-\Delta
\Gamma}{2}} \cos \Delta mt + e^{-\Delta \Gamma t}\right\}
\end{equation}
where $\Delta m=m_2-m_1$, $\Delta\Gamma=\Gamma_2-\Gamma_1$ are the mass and width
differences of the mass eigenstates and $\Gamma=\frac{\Gamma_1+\Gamma_2}{2}$. In the limit 
$\Delta m,\Delta\Gamma <<\Gamma $, 
mixing effects are parametrized by two dimensionless amplitudes $x=\frac{\Delta
m}{\Gamma}$ and $y=\frac{\Delta\Gamma}{2\Gamma}$. The experimental measurement is
performed by analyzing wrong sign semileptonic and hadronic decays; the presence
of doubly Cabibbo-suppressed decays complicates the measurement in the hadronic
sector. Assuming CP conservation, the wrong sign to right sign  decay rate is
\begin{equation}
r_D(t) = e^{-\Gamma t } \left( R_{DCS} + \sqrt{R_{DCS}}\,y'\Gamma t
+\frac{1}{4}(x'^2+y'^2)\Gamma^2 t^2\right)
\end{equation}
where $y'=y\cos\delta-x\sin\delta$ and $x'=x\cos\delta+y\sin\delta$, $\delta$
being the strong phase difference between the Cabibbo-favored and Doubly
Cabibbo-suppressed Decay (DCS). The three terms come from DCS decays,
interference, and mixing. In the semileptonic mixing only the last term
matters. The up-to-date experimental situation is summarized in
Fig.~\ref{fig:mix04} \cite{cleo_first,cicerone}.
In the semileptonic sector, FOCUS has measured, preliminarily, $r_D$ in the mode
$D^0\to K^+\mu^-\bar\nu_{\mu}$ to be less than 0.131\% @ 95\%\,C.L.; CLEO-c
expects $r_D$ less than $10^{-4}$\% @ 95\%\,C.L.; SuperBaBar estimates that, with
10\,ab$^{-1}$, the sensitivity would be $r_D<5 \times 10^{-4}$. With $10^9$
reconstructed charm decays BTeV expects to reach a sensitivity of 1--$2\times
10^{-5}$. An observation of mixing at the level of $10^{-4}-10^{-5}$, which could
be probed by a high-sensitivity experiment, would be a signal of New Physics. In
Fig.~\ref{fig:mix04} measurements of the lifetime difference between $D^0$ decays
to CP-even and CP-odd final states are also shown. If CP violation in neutral
$D$-meson decays is negligible we can write
\begin{equation}
y_{CP} = \frac{\Gamma(\mathrm{CP\;even})-\Gamma(\mathrm{CP\;odd})}
{\Gamma(\mathrm{CP\;even})+\Gamma(\mathrm{CP\;odd})} \sim \frac{\Gamma(D^0\to
K^+K^-)}{\Gamma(D^0\to K^-\pi^+)} -1
\end{equation}
The theoretical prediction spread of the mixing parameters is wild, as shown in
the right side of Fig.~\ref{fig:mix04}; independently of New Physics discovery,
accurate experimental measurements are necessary.

\subsection{CP Violation}

CP violation can occur in charm decays via the interference of two weak decay
amplitudes to the same final state. \emph{Indirect} CP violation is mediated by
$D^0$--$\bar D^0$ mixing. The interference is the largest if the two amplitudes
are roughly equal. However, since mixing  is expected to be small, indirect CP
violation is not expected to be a big effect. In absence of mixing, a decay
mode (e.g., a Cabibbo-suppressed decay) that has two weak amplitudes
contributing to the same final state can exhibit \emph{direct} CP violation.
Final State Interactions (FSI) can induce a phase shift between the two weak
amplitudes, leading to a decay-rate asymmetry between a charm meson decay and
its CP conjugate: $\Gamma(D\to f) \ne \Gamma(\bar D\to\bar f)$. FSI are
substantial in charm decay, and the two weak amplitudes can be similar in size.
Asymmetries as large as 0.1--1\% are possible in the Standard Model. In
Table~\ref{tab:cpv} the experimental scenario is reported.

\begin{table}[htbp]
  \def\M{\hphantom-}
  \centering
  \caption{Measurements of CP asymmetries.}
  \label{tab:cpv}
  \begin{displaymath}
  \begin{array}{lcc}
  \hline \hlinestrut
  $Decay mode$ & $E791$ & $CLEO$ \\
  D^0\to K^-K^+ & -0.010 \pm 0.049 \pm 0.012 & \M0.000 \pm 0.022 \pm
  0.008 \\
  D^0\to\pi^-\pi^+ & -0.049 \pm 0.078 \pm 0.030 & \M0.030 \pm 0.032
  \pm 0.008 \\
  D^0\to K_s K_s & & -0.23 \pm 0.19  \\
  D^0\to K_s\pi^0 & & \M0.001 \pm 0.013  \\
  D^0\to\pi^0\pi^0 & & \M0.001 \pm 0.048  \\
  D^+\to K^-K^+\pi^+ & -0.014 \pm 0.029 & \\
  D^+\to\pi^-\pi^+\pi^+ & -0.017 \pm 0.042 & \\[1ex]
  & $FOCUS$ & $CDF$ \\
 D^0\to K^-K^+    &  -0.001 \pm 0.022 \pm 0.015 & \M0.020 \pm 0.012\pm 0.006 \\
 D^0\to\pi^-\pi^+ & \M0.048 \pm 0.039 \pm 0.025 & \M0.030 \pm 0.013\pm 0.006 \\
  D^+\to K^-K^+ \pi^+ & \M0.006 \pm 0.011\pm 0.005   &    \\
  D^+\to K_s\pi^+ & -0.016 \pm 0.015 \pm 0.009  &  \\
  D^+\to K_s K^+ & \M0.071 \pm 0.061 \pm 0.012  &  \\
  \hline
  \end{array}
\end{displaymath}
\end{table}

Expectations come from CLEO-c, which aims at exploiting the quantum coherence
of the $D^0$--$ \bar D^0$ pair produced. The process
$e^+e^-\to\psi''\to D^0\bar D^0$
produces a CP+ eigenstate in the first step, since $\psi''$ has
$J^{PC}=1^{--}$. Consider the case where both the $D^0$ and the $\bar D^0$
decay into CP eigenstates. Then the decays
$\psi^{''}\to f^i_+ f^i_+$ \, or $f^i_- f^i_-$
are forbidden, where $f_+/f_-$ denotes a CP+ / CP- eigenstates. This is because
$CP( f^i_ {\pm}f^i_{\pm}) = (-1)^l=-1$
for the l=1 $\psi^{''}$. Thus, observation of a final state such as
$(K^+K^-)(\pi^+\pi^-)$ constitutes evidence of CP violation. Of course this method
requires a complete understanding of the Initial State Radiation (ISR) effects. CP
asymmetries measured so far in the charm sector are consistent with zero within
the errors. The high statistics soon available from experiments such as Belle,
BaBar and CLEO-c would be able to probe CP asymmetries at the $10^{-3}$ level.
BTeV and will extend the sensitivity below the $10^{-3}$ level.

\subsubsection{CP investigation via Dalitz-plot analysis}
\label{sec:CPV}

\begin{figure}[hbtp]
  \centering
  \epsfig{file=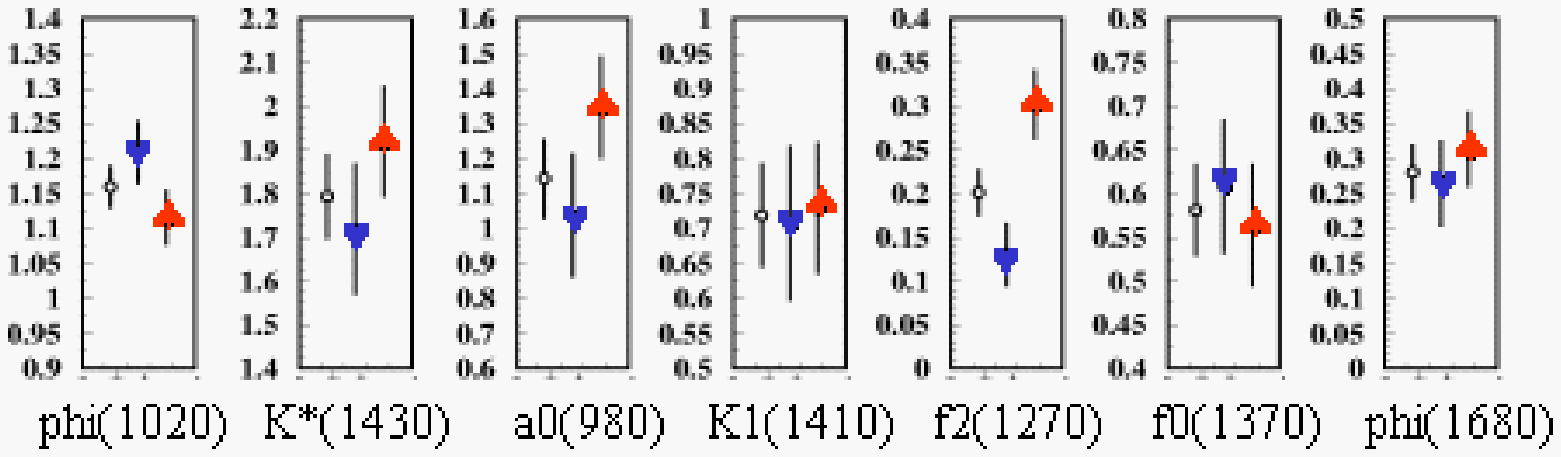,width=0.49\linewidth}
  \epsfig{file=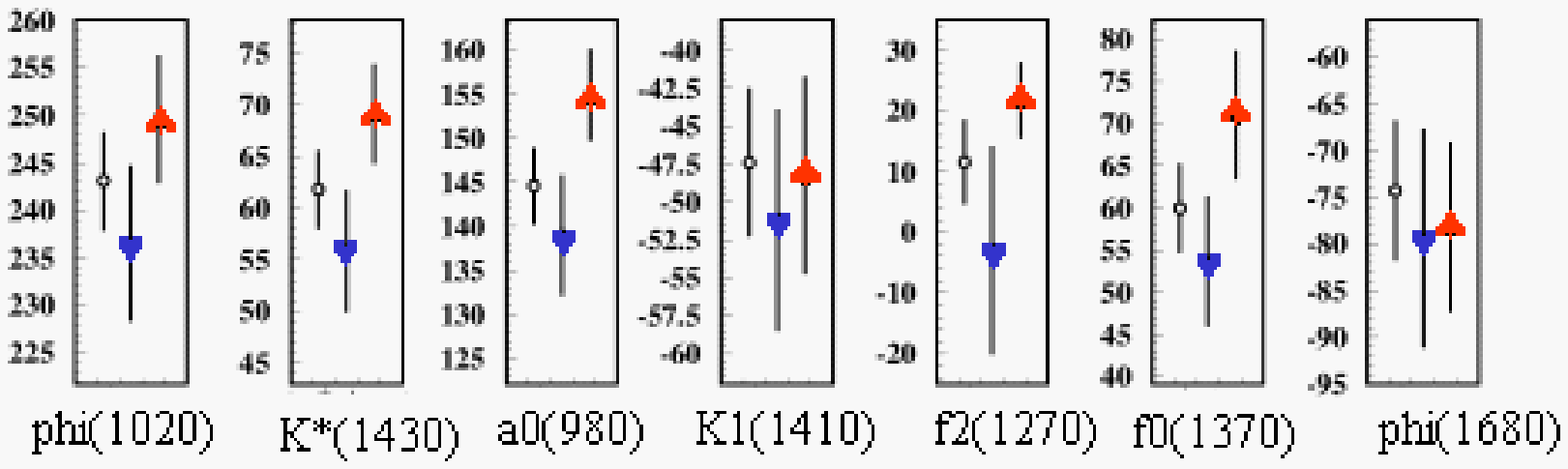,width=0.49\linewidth}
  \caption{Amplitude coefficient and relative phases (in degrees) of all the
    resonant contributions of the $D^+\to K^-K^+\pi^+$ decay: the circles are
    for the full sample, upward triangles for $D^-$ and downward triangles for
    $D^+$.}
  \label{fig:CPV}
\end{figure}

Dalitz-plot analysis has recently emerged as a unique tool to investigate CPV
effects. Amplitude analysis gives the full observation of a three-body decay,
providing access to the coefficients and phases of the various channels. CP
effects are intimately connected to phase phenomena. The comparison between the
relative phases of the CP-conjugate states for the various resonant channels can
address CPV effects. FOCUS analyzed the $D^+\to K^-K^+\pi^+$ channel
\cite{ichep2002} and CLEO the $D^0\to K_S\pi^+\pi^-$ channel \cite{asner}. In
general each measured phase in the Dalitz plot can be interpreted as a sum of two
components, one CP conserving, the other CP violating. Under CP conjugation the
first does not switch sign, while the second does. The $D^+\to K^-K^+\pi^+$
channel is a good candidate to observe CP effects. This decay is Cabibbo
suppressed and thus contains two diagrams contributing to the final state,
spectators and penguins; it is also dominated by strong effects, as discussed
above, and thus satisfies the two necessary requirements for detecting a CP
asymmetry. FOCUS \cite{ichep2002} has performed a new Dalitz-plot study, which
proceeds through a complete Dalitz analysis of the $D^+/D^-$ split samples. The
preliminary results on $D^+/D^-$ coefficients and phases, shown in
Fig.~\ref{fig:CPV}, do not point to any CPV manifestation. The relative phases are
measured with an accuracy of about 5--10$^\circ$. The errors are affected by a
large correlation among the various resonant components necessary to fit the
Dalitz plot. A more robust formalism, such as the \emph{K-matrix} approach, which will be 
discussed later in this paper, might
allow for a good fit with fewer resonances, thus ameliorating the accuracy of the
phase measurements. At any rate a factor of 100 more statistics, as hopefully
available in BTeV, will allow for reaching sensitivities of a degree.

\subsection{Rare decays}

\begin{figure}[hbtp]
  \centering
  \epsfig{file=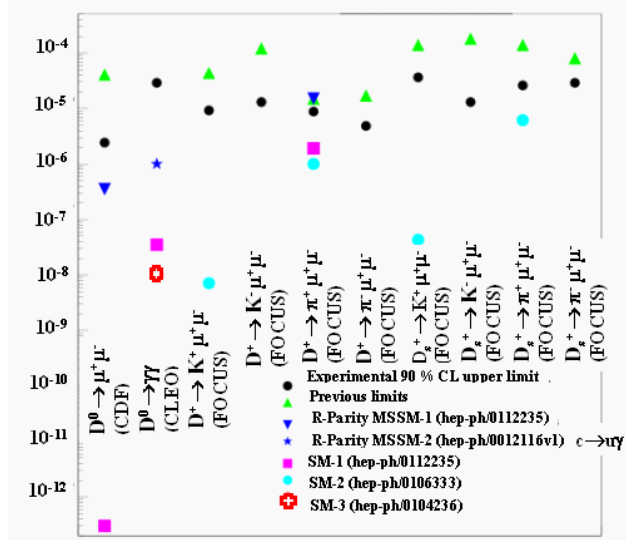,width=0.52\linewidth}
  \caption{Measurements of rare decays.}
  \label{fig:rare04}
\end{figure}

The search for rare and forbidden decays is enticing since Standard Model
predictions tend to be beyond the reach of current experiments, and a signal is a
sign of unexpected physics. Standard Model predictions of rare decays are
dominated by long-range effects, which are notoriously difficult to calculate and
experimental measurements are thus crucial. The current experimental scenario
concerning rare decay branching ratios is reported in Fig.~\ref{fig:rare04}. FOCUS
\cite{rare} has improved many of the previous limits of a factor from 2 to 14. The
strength of the FOCUS fixed target experiment is the excellent vertexing, which
was used to require that the two candidate leptons form a good, well separated
vertex from the primary. CDF and D0 have efficient di-lepton triggers and future
measurements are promising. The reported CDF result is obtained only with a
statistics of 65\,pb$^{-1}$. A new limit of $2 \times 10^{-6}$ has been recently
provided by the Hera-B experiment\cite{herab}. CLEO-c sensitivity is estimated at
the level of $10^{-6}$.

\section{Standard Model measurements and lessons for the future}

The variety of measurements above briefly discussed represents an excellent
opportunity for searching New Physics in the charm sector. Effects beyond the
Standard Model are, as already pointed out, expected to be small, making their
investigations really challenging. Possible tiny inconsistencies and deviations
from expected behavior can be interpreted as sign of New Physics only if the
Standard Model phenomenology is correctly parametrized in the analysis
procedure. The main lesson coming from the mature field of charm is that the
advantages of high statistics to interpret the Heavy Flavor decay dynamics will
vanish in the absence of a strategy to control strong effects among particles
involved in weak decay processes. A few examples from the FOCUS experiment will
be discussed.

\subsection{Unexpected new results in the semileptonic sector}

\begin{figure}[hbtp]
  \centering
  \epsfig{file=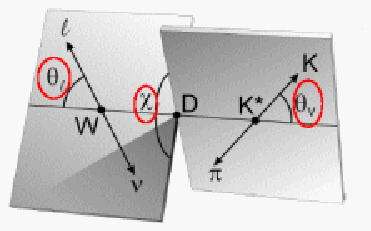,width=0.3\linewidth}
  \epsfig{file=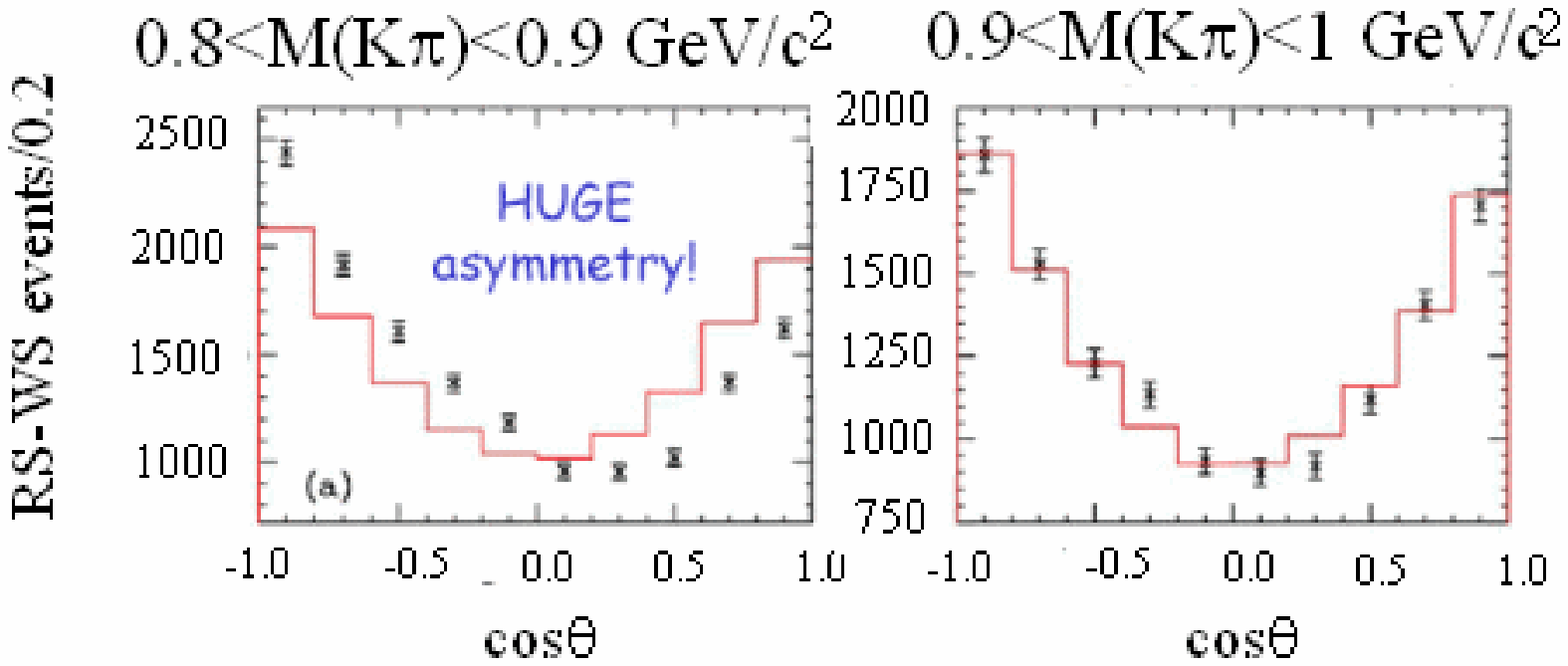,width=0.5\linewidth}
  \caption{The definition of the
  kinematic variables for the $D^+\to\bar{K}^{*0}\mu^+\nu$ and $\cos\theta_V$
  distribution.}
  \label{fig:kinematics}
\end{figure}

Semileptonic decays have always been considered the best candidates to study the
charm phenomenology. Decay rates can be calculated from first principles, e.g,
Feynman diagrams; due to the presence of only one hadron in the final state we do
not have to worry much about final state interaction complications; the hadronic
part of the decay can be contained in proper form factors, which, on their turn
can be predicted by theory such as HQET, LGT and Quark models. However recent
FOCUS studies have shown that decays in the semileptonic sector also reveal the
presence of quantum mechanical effects and hadronic complications have to be dealt
with in the analysis. Of particular relevance is the study of $D^+\to
K^-\pi^+\mu^+\nu$ \cite{slint}. The FOCUS mass spectrum appears, as expected,
dominated by $\bar{K}^{*0}(892)$; yet an unexpected forward-backward asymmetry was
exposed in the $\cos\theta_V$ variable, defined in Fig.~\ref{fig:kinematics}. The
asymmetry is striking below the pole and essentially absent above the pole: it
thus suggests quantum-mechanical interference between the Breit--Wigner amplitude
describing the $\bar{K}^{*0}(892)$ and a broad or nearly constant $S$-wave
amplitude. FOCUS has tried to describe this behavior of the decay distribution for
$D^+\to K^-\pi^+\mu^+\nu$ through a simple model. It has been written in terms of
the three helicity-basis form factors: $H_+,H_0,H_-$, and the $S$-wave amplitude
modeled as a constant, with modulus $A$ and phase $\delta$ has been added. Angular
momentum conservation restricts the $S$-wave contribution to the $H_0$ piece that
describes the amplitude for having the virtual $W^+$ in a zero helicity state
relative to its momentum vector in the $D^+$ rest-frame.

\begin{figure}[hbtb]
  \centering
  \epsfig{file=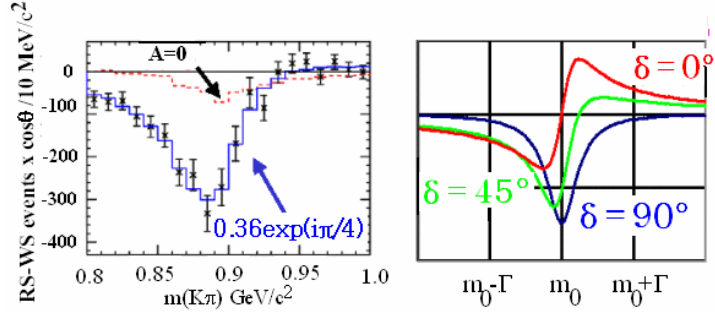,width=0.6\linewidth}
  \caption{Asymmetry distribution in $K\pi$ invariant mass. The dashed line is
  the MC with no $S$-wave amplitude, the solid line is the MC with an $S$-wave
  amplitude of $0.36\exp(i\pi/4)\,$(GeV)$^{-1}$. Different modelings of the
  $S$-wave are also shown.}
  \label{fig:masskpi}
\end{figure}

The shape of the $\cos\theta_V$ term versus $m_{K\pi}$ turns out to be a strong
function of the interfering $S$-wave amplitude phase $\delta$.
Figure~\ref{fig:masskpi} shows how the $\cos\theta_V$ distribution (properly
weighted) is consistent with a constant $S$-wave amplitude of the form
$0.36\exp(i\pi/4)\,$(GeV)$^{-1}$, although, at this stage of the analysis, the
solution is not unique; alternative modelings of the $S$-wave amplitude, as
shown in the same Fig.~\ref{fig:masskpi}, could also fit the data. 
The measured phase of $\pi/4 $ is consistent with that found by LASS  from the
$K \pi$ phase-shift analysis. The
hypothesis of the broad scalar resonance $\kappa$ is disfavored since it would
imply a phase-shift of approximately $90^{\circ}$ with respect to
$\bar{K}^{*0}(892)$, unlikely to manifest in the semileptonic sector where FSI
are not expected to be wild \cite{doris}.
A more precise determination can be obtained through a complete analysis of the
form factor ratios $r_v$ and $r_2$ \cite{slff}. FOCUS has found $r_v =1.504 \pm
0.057 \pm 0.039$ and $r_2 = 0.875 \pm 0.049 \pm 0.064$. For the $S$-wave we
obtained an amplitude modulus of $A=0.330\pm0.022\pm0.015\,$GeV$^{-1}$ and a phase
$\delta =0.68 \pm 0.07 \pm 0.05\,$rad, in reasonable agreement with the very
informal previous estimate of $A=0.36\exp(i\pi/4)\,$GeV$^{-1}$. The inclusion of
the $S$-wave amplitude has dramatically improved the quality of the form factor
fits. A measurement of the branching ratio $\frac{\Gamma(D^+\to\bar
K^{*0}\mu^+\nu)}{\Gamma(D^+\to
K^-\pi^+\pi^+)}=0.602\pm0.01(\mbox{stat.})\pm0.021(\mbox{syst.})$ has then been
performed including interference with this broad scalar resonance.
To complete the semileptonic study, FOCUS has performed an analysis of the
$D_s^+\to\phi\mu^+\nu$ final state; the
$\frac{\Gamma(D_s^+\to\phi\mu^+\nu)}{\Gamma(D_s^+\to\phi\pi^+)}$ has been measured
to be $0.54 \pm 0.033(\mbox{stat.}) \pm 0.048 (\mbox{syst.})$ \cite{slbr}, and a
new measurement of form factor ratios has been recently published~\cite{dsff}.

\subsection{\boldmath Unexpected complications in hadronic decay dynamics:
$D^+$ and $D^+_s\to\pi^+\pi^-\pi^+$}

Charm-meson decay dynamics has been extensively studied in the last decade. The
analysis of the three-body final state by fitting Dalitz plots has proved to be a
powerful tool for investigating effects of resonant substructure, interference
patterns, and final state interactions in the charm sector. An example of
Dalitz-plot analysis to study CP-violation effects has been already discussed in a
section of this paper. The isobar formalism, which has traditionally been applied
to charm amplitude analyses, represents the decay amplitude as a sum of
relativistic Breit--Wigner propagators multiplied by form factors plus a term
describing the angular distribution of the two body decay of each intermediate
state of a given spin. Many amplitude analyses require detailed knowledge of the
light-meson sector. In particular, the need to model intermediate scalar particles
contributing to the charm meson decays into three-body hadronic channels has caused
experimentalists of the field to question the validity of the Breit--Wigner
approximation for the description of the relevant scalar resonances
\cite{pdg_rev,penn1,eef}. Resonances are associated with poles of the \emph{S-matrix}
in the complex energy plane. The position of the pole in the complex energy plane
provides the fundamental, model-independent, process-independent resonance
description. A simple Breit--Wigner amplitude corresponds to the most elementary
type of extrapolation from the physical region to an unphysical-sheet pole. In the
case of a narrow, isolated resonance, there is a close connection between the
position of the pole on the unphysical sheet and the peak we observe in
experiments at real values of the energy. However, when a resonance is broad and
overlaps with other resonances, then this connection is lost.  The Breit--Wigner
parameters measured on the real axis (mass and width) can be connected to the
pole-positions in the complex energy plane only through models of analytic
continuation.
A formalism for studying overlapping and many channel resonances has been proposed
long ago and is based on the  \emph{K-matrix} \cite{wigner,chung} parametrization.
This formalism, originating in the context of two-body scattering, can be
generalized to cover the case of production of resonances in more complex
reactions \cite{aitch}, with the assumption that the two-body system in the final
state is an isolated one and that the two particles do not simultaneously interact
with the rest of the final state in the production process \cite{chung}. The
\emph{K-matrix} approach allows for including the positions of the poles in the
complex plane directly in the analysis, incorporating in the charm
analysis the results from light spectroscopy experiments \cite{penn2,anisar1}. In
addition, the \emph{K-matrix} formalism provides a direct way of imposing the
two-body unitarity constraint which is not explicitly guaranteed in the simple
isobar model. Minor unitarity violations are expected for narrow, isolated
resonances but more severe ones exist for broad, overlapping states. The validity
of the assumed quasi two-body nature of the process of the \emph{K-matrix}
approach can only be verified by a direct comparison of the model predictions with
data. In particular, the failure to reproduce three-body-decay features would be a
strong indication of the presence of neglected three-body effects.

\subsection{The isobar formalism}

\begin{figure}[htb]
  \setlength{\unitlength}{0.6mm}
  \centering
  \begin{picture}(110,50)(0,0)
    \thicklines
    \put(30,25){\circle{12}}
    \put(27,24){$F_D$}
    \put(70,25){\circle{12}}
    \put(68,24){$F_r$}
    \put(-15,24){$(p_\pi+p_D)_\mu$}
    \put(78,24){$(p_{\pi^+}-p_{\pi^-})_\nu$}
    \put(48,28){$\rho$}
    \put(25,-2){\makebox(50,20){$\displaystyle\frac
                                 {g^{\mu\nu}-q^\mu q^\nu/m_0^2}
                                 {q^2-(m_0-i\Gamma/2)^2}$}}
    \put(04,47){$\pi$}
    \put(93,47){$\pi^+$}
    \put(04,00){$D$}
    \put(93,00){$\pi^-$}
    \put(27,28){\vector(-1,1){10}}
    \put(73,28){\vector(1,1){10}}
    \put(18,37){\line(-1,1){10}}
    \put(82,37){\line(1,1){10}}
    \put(27,22){\line(-1,-1){10}}
    \put(73,22){\line(1,-1){10}}
    \put(08,03){\vector(1,1){10}}
    \put(92,03){\vector(-1,1){10}}
    \multiput(35,25)(4.8,0){7}{\line(1,0){2.0}}
  \end{picture}
  \caption{The $D^+\to\pi\pi\pi$ decay diagram.}
  \label{fig:fey}
\end{figure}
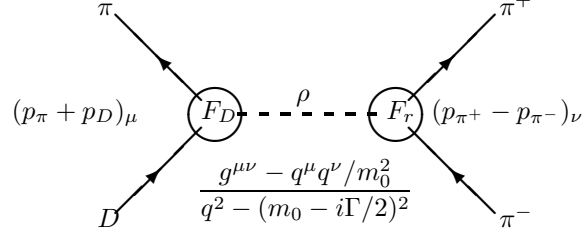

The formalism traditionally applied to three-body charm decays relies on the
so-called isobar model. A resonant amplitude for a quasi-two-body channel, of
the type
\begin{equation}
\label{reso}
  \begin{array}{cl}
    D\to &  r + c     \\[-0.5ex]
          & \spig a + b
  \, ,
  \end{array}
\end{equation}
is interpreted \emph{\`{a} la} Feynman. For the decay $D\to\pi\pi\pi$ of
Fig.~\ref{fig:fey},
a $D\to\pi$ current with form factor $F_D$ interacts with a di-pion current
with form factor $F_r$ through an unstable propagator with an imaginary width
contribution in the propagator mass. Each resonant decay function is thus,
\begin{equation}
  A =
  F_DF_r
  \times
  |\bar c|^J |\bar a|^J P_J(\cos\Theta^r_{ac})
  \times
  BW(m_{ab})
  \,
  \label{twobody}
\end{equation}
i.e., the product of two vertex form factors (Blatt--Weisskopf
momentum-dependent factors), a Legendre polynomial of order $J$ representing
the angular decay wave function, and a relativistic Breit--Wigner (BW). In this
approach, already applied in the previous analyses of the same channels
\cite{e687_dpds}, the total amplitude (Eq.~\ref{totamp}) is assumed to consist
of a constant term describing the direct non-resonant three-body decay and a
sum of functions (Eq.~\ref{twobody}) representing intermediate two-body
resonances.

\begin{equation}
  A(D) = a_0 e^{i\delta_0} + \sum_i a_i e^{i\delta_i} A_i \, ,
   \label{totamp}
\end{equation}

\subsection{The \boldmath{$K$}-\emph{matrix} formalism}

For a well-defined wave of specific isospin and spin \emph{IJ}, characterized
by narrow and isolated resonances the propagator is, as anticipated, of the
simple BW form. In contrast, when the specific wave \emph{IJ} is characterized
by large and heavily overlapping resonances, just as the scalars, the
propagation is no longer dominated by a single resonance, but is the result of
complicated interplay among the various resonances. In this case, it can be
demonstrated on very general grounds that the propagator may be written in the
context of the \emph{K-matrix} approach as
\begin{equation}
(I -iK \cdot \rho)^{-1}
 \label{eq_prop}
\end{equation}
where \emph{K} is the matrix for the scattering of particle $a$ and $b$ ( Eq. \ref{reso}) and $\rho$
is the phase-space matrix. In this picture, the production process is viewed as
consisting of an initial preparation of several states, which then propagate via
the term $(I-iK\rho)^{-1}$ into the final state. In particular, the three-pion
final state can be fed by an initial formation of $(\pi\pi)\pi$, $(K\anti K)\pi$,
$(\eta\eta) \pi$, $(\eta\eta') \pi$ and multi-meson states (mainly four-pion
states at $\sqrt{s}<1.6$\,GeV). While the need for a \emph{K-matrix}
parametrization, or in general for a more accurate description than the isobar
model, might be questionable for the vector and tensor amplitudes, since the resonances
are relatively narrow and well isolated, this parametrization is needed for the
correct treatment of scalar amplitudes. Indeed the $\pi\pi$ scalar resonances are
large and overlap each other in such a way that it is impossible to single out the
effect of any one of them on the real axis. In order to write down the propagator,
we need the scattering matrix. To perform a meaningful fit to $D$ mesons to
three-pion data, a full description of the scalar resonances in the relevant
energy range, updated to the most recent measurements in this sector is needed. At
the present time the only self-consistent description of $S$-wave isoscalar
scattering is that given in the \emph{K-matrix} representation by Anisovich and
Sarantsev in \cite{anisar1} through a global fit of the available scattering
data from the $\pi\pi$ threshold up to 1900\,MeV. FOCUS has performed the first
fit to charm data with the \emph{K-matrix} formalism \cite{kmatrix} in the $D^+$
and $D^+_s\to\pi^+\pi^-\pi^+$ channels. The \emph{K-matrix} used is that of
\cite{anisar1}:
\begin{equation}
 K_{ij}^{00}(s) =
  \left\{
    \sum_\alpha \frac{g^{(\alpha)}_i g^{(\alpha)}_j}{m^2_{\alpha}-s}
 +  f^\mathrm{scatt}_{ij}\frac{1\,\mathrm{GeV}^2
 - s_0^\mathrm{scatt}}{s-s_0^\mathrm{scatt}}
  \right\}
 \times \frac{s-s_A m^2_{\pi}/2}{(s-s_{A0})(1-s_{A0})}.
  \label{eq_sarantsev}
\end{equation}
The factor $g^{(\alpha)}_i$ is the coupling constant of the \emph{K-matrix}
pole $\alpha$ to meson channel $i$; the parameters $f^\mathrm{scatt}_{ij}$ and
$s_0^\mathrm{scatt}$ describe a smooth part of the \emph{K-matrix} elements;
the factor $\frac{s-s_A m^2_{\pi}/2}{(s-s_{A0})(1-s_{A0})}$ suppresses a false
kinematical singularity in the physical region near the $\pi\pi$ threshold
(Adler zero). The \emph{K-matrix} values of \cite{anisar1} generate a physical
\emph{T-matrix}, \mbox{$T=(I-i\rho\cdot K)^{-1}K$}, which describes the
scattering in the $(00)^{++}$-wave with five poles, whose masses, half-widths,
in GeV are (1.019,\,0.038), (1.306,\,0.167), (1.470,\,0.960), (1.489,\,0.058)
and (1.749,\,0.165).
The  \emph{K-matrix} formalism, originated in the context of the two-body scattering, can
be generalized to deal with formation of resonances in more complex reactions,
through the $P$-\emph{vector} \cite{aitch} approach. The decay amplitude for the
$D$ meson into three-pion final state, where $\pi^+\pi^-$ are in a
$(IJ^{PC}=00^{++}$)-wave, can thus written as

\begin{eqnarray}
  A(D\to (\pi^+\pi^-)_{00^{++}}\pi^+)= F_1 =
  (I-iK\rho)^{-1}_{1j} \qquad&\qquad\nonumber \\
  \times \left\{
    \sum_\alpha \frac{\beta_{\alpha}g_{j}^{(\alpha)}}{m^2_{\alpha}-m^2}
  + f_{1j}^\mathrm{prod}\frac{1\,\mathrm{GeV}^2
  - s_0^\mathrm{prod}}{s-s_0^\mathrm{prod}}
  \right\}
  \times \frac{s-s_A  m^2_{\pi}/2}{(s-s_{A0})(1-s_{A0})}.
  \label{eq_kmat}
\end{eqnarray}
where $\beta_{\alpha}$ is the coupling to the pole $\alpha$ in the `initial'
production process, $f_{1j}^\mathrm{prod}$ and $s_0^\mathrm{prod}$ are the
$P$-\emph{vector} slowly varying (SVP) parameters. In the end, the complete decay
amplitude of the $D$ meson into three-pion final state is \cite{kmatrix}:
\begin{equation}
  A(D) = a_0 e^{i\delta_0} + \sum_i a_i e^{i\delta_i} A_i + F_1
  \label{totampK}
\end{equation}
where the index $i$ now runs only over the vector and tensor resonances, which
can be safely treated as simple Breit--Wigner's (see Eq.~\ref{twobody}). In the
fit to the data, the \emph{K-matrix} parameters are fixed to the values of
\cite{anisar1}, which consistently reproduce measured $S$-wave isoscalar
scattering. The free parameters are those peculiar to the $P$-\emph{vector},
i.e., $\beta_{\alpha}$, $f^\mathrm{prod}_{1j}$ and $s_0^\mathrm{prod}$, and
those in the remaining isobar part of the amplitude, $a_i$ and $\delta_i$.

\begin{figure}[htbp]
  \centering
  \subfigure[$D_s^+$]
  {%
    \epsfig{file=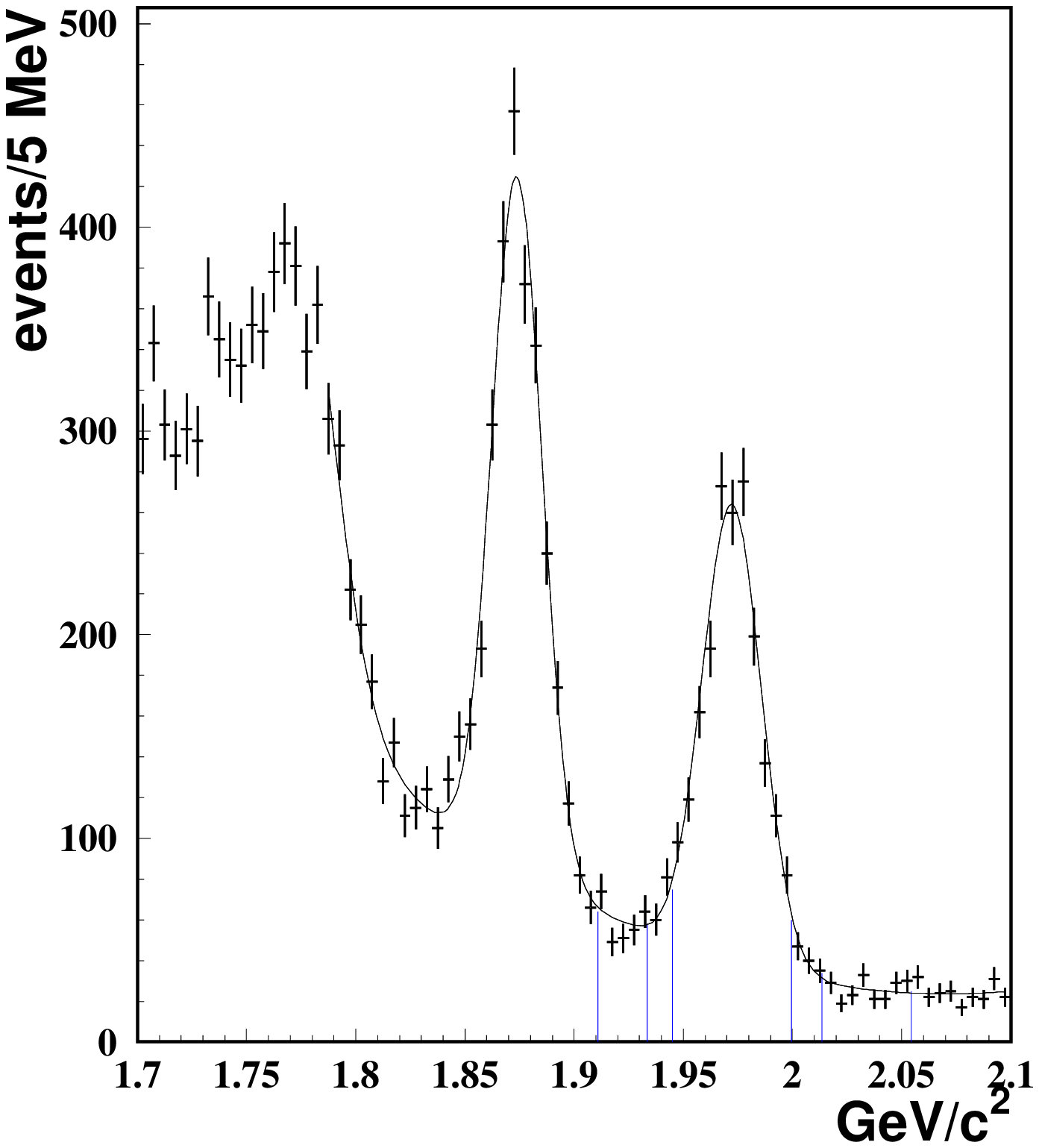,angle=-90,width=0.27\linewidth}
  }
  \subfigure[$D^+$]
  {%
    \epsfig{file=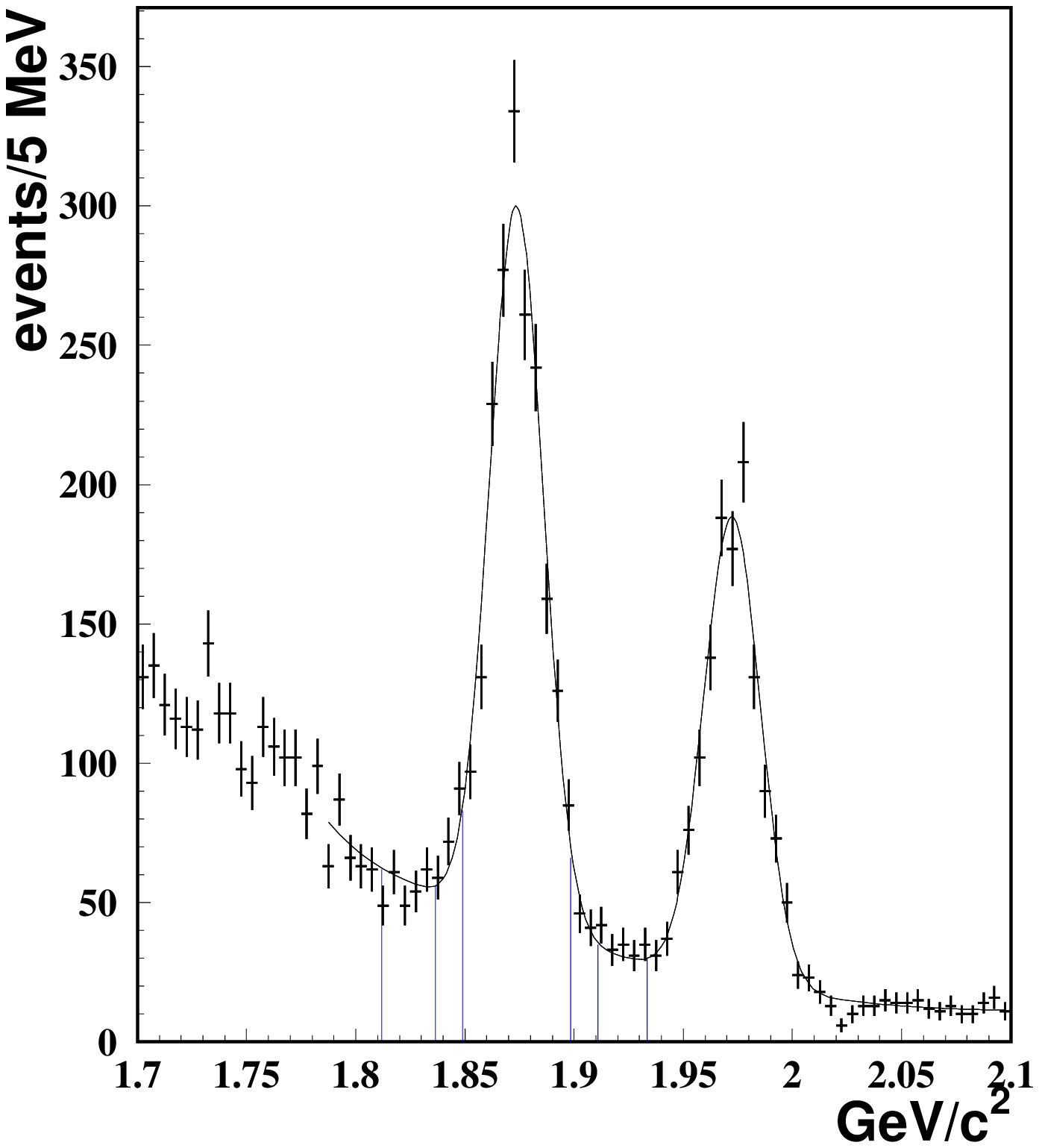,angle=-90,width=0.27\linewidth}
  }
  \caption{Signal and side-band regions of the three-pion invariant-mass
  distribution
  for a) $D_s^+$ and b) $D^+$ Dalitz-plot analysis respectively.}
  \label{mass}
\end{figure}

\begin{figure}[htbp]
  \begin{center}
  \subfigure[$D_s^+$]
  {%
    \epsfig{file=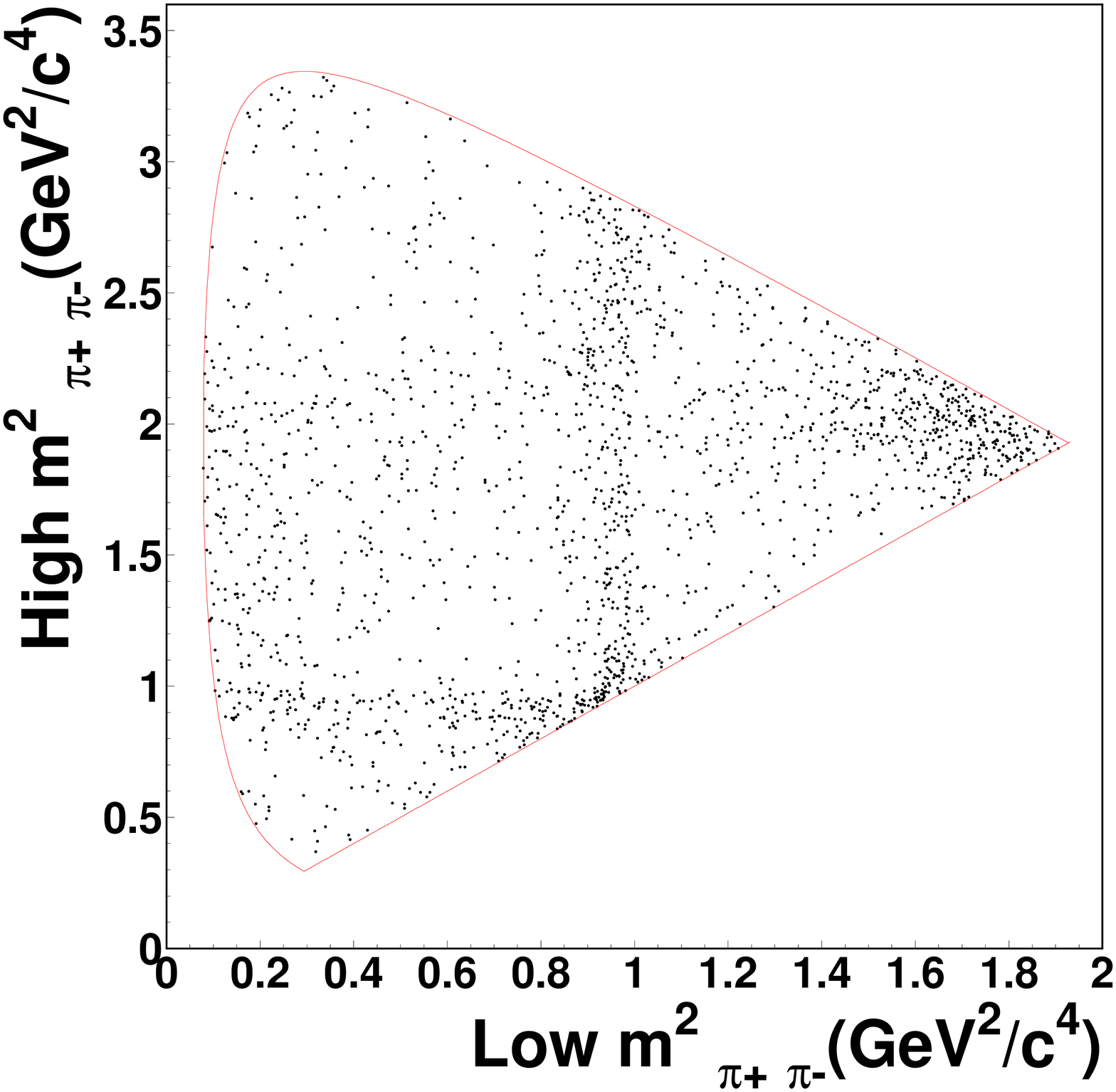,width=0.3 \linewidth}
  }
  \subfigure[$D^+$]
  {%
    \epsfig{file=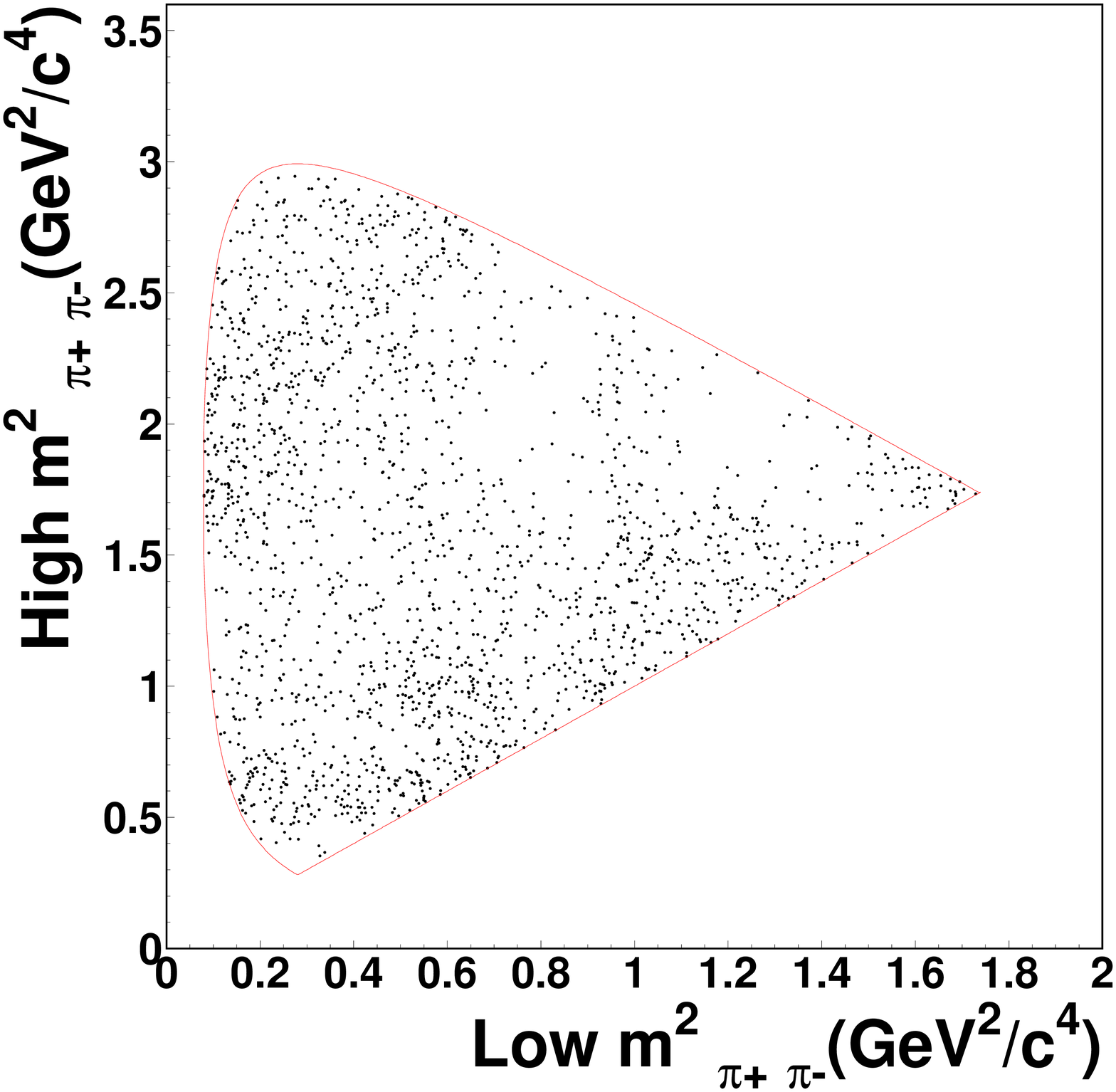,width=0.3\linewidth}
  }
  \end{center}
  \caption{a) $D_s^+$ and b) $D^+$ Dalitz plots.}
  \label{dalitz}
\end{figure}

The three-pion selected samples (Fig.~\ref{mass}) consist of $1527\pm51$ and
$1475\pm50$ events for the $D^+$ and $D_s$ respectively. The Dalitz-plot
(Fig.~\ref{dalitz}) analyses are performed on yields within $\pm 2\sigma$ of
the fitted mass value.

\subsubsection{\boldmath Results for the $D^+_s\to\pi^+\pi^-\pi^+$ decay}

\begin{figure}[htbp]
 \centering
 \subfigure
  {%
    \epsfig{file=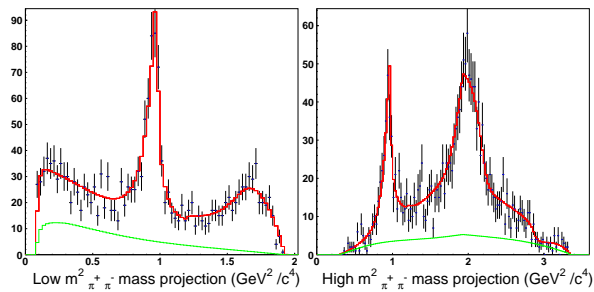,width=0.5\linewidth}
  }
 \caption{$D_s^+$ Dalitz-plot projections with the \emph{K-matrix} fit
 superimposed. The background shape under the signal is also shown.}
 \label{ds_proj_Kmatrix}
\end{figure}

The general procedure adopted for the fits consists of several
successive steps in order to eliminate contributions whose effects on fits are
marginal.  Initially all the well established, non-scalar resonances decaying to
$\pi^+\pi^-$ with a sizeable branching ratio are considered. Contributions are
removed if their amplitude coefficients, $a_i$ of Eq.~\ref{totampK}, are less than
$2\,\sigma$ significant \emph{and} the fit confidence level increases due to the
decreased number of degrees of freedom in the fit. The \emph{P-vector} initial
form includes the complete set of \emph{K-matrix} poles and slowly varying
function (SVP) as given in reference \cite{anisar1}; $\beta_{\alpha}$ as well as
the $f^{\mathrm{prod}}_{1j}$ terms of Eq.~\ref{eq_kmat} are removed with the same
criteria. The fit confidence levels (C.L.) are evaluated with a $\chi^2$ estimator
over a Dalitz plot with bin size adaptively chosen to maintain a minimum number of
events in each bin. Once the minimal set of parameters is reached, addition of
each single contribution previously eliminated is reinstated to verify that the
C.L. does not improve.
The resulting fit fractions \footnote{The quoted fit fractions are defined as the
ratio between the intensity for a single amplitude integrated over the Dalitz plot
and that of the total amplitude with all the modes and interferences present.},
phases and amplitude coefficients are quoted in Table~\ref{table_ds_kmatrix}. 
Both the three-body non-resonant and $\rho^0(770)\pi^+$ components are 
not required by the fit. This result is to be compared with that obtained with the
simple isobar model, which requires a  non-resonant component of about 25\%  to
get a decent fit to the data \cite{scawork}. This component, which crosses the Dalitz plot
uniformly, seems to compensate, with its interference with the other contributions, 
for the inability of the model to properly describe
some non-trivial resonant features not properly accounted for in the model. 
In this way the potentiality of the
Dalitz-plot analysis to gauge the level of the annihilation contribution in the
charm hadronic decays is limited. An additional difficulty with the isobar model 
is the general poor knowledge of scalar resonances: the measurements 
reported in the PDG are dispersed over a wide range of values and can not be used as input
parameters of charm decay amplitudes. Masses and widths of the corresponding Breit-Wigner forms
have to be let free in the fit:  
the isobar model can thus be viewed as an
effective model able to reproduce the data with a sum of effective resonances but 
its phenomenological interpretation has to be considered with caution.
The entire $S$-wave
contribution obtained with the \emph{K-matrix} formalism is represented by a 
single fit fraction since, as previously
discussed, one cannot distinguish the different resonance or SVP $S$-wave
contributions on the real axis. The $D_s^+$ Dalitz projections of FOCUS data are
shown in Fig.~\ref{ds_proj_Kmatrix} superimposed with final fit projections. The
fit C.L. is 3\%.
\begin{table}[htbp]
  \centering
  \caption{Fit results from the \emph{K-matrix} model for $D_s^+$.}
  \label{table_ds_kmatrix}
  \begin{tabular}{cccc}
  \hline \hlinestrut
    decay channel  & fit fraction (\%) & phase (deg) & amplitude coefficient \\
  \hline \hlinestrut
      ($S$-wave)\,$\pi^+$     &  $87.04 \pm 5.60 \pm 4.17$ & 0 (fixed)
       & 1 (fixed)  \\
  $f_2(1270)\,\pi^+$    &  $9.74 \pm 4.49 \pm 2.63$&  $168.0 \pm 18.7 \pm 2.5$
   & $0.165 \pm 0.033\pm 0.032$ \\
  $\rho^0(1450)\,\pi^+$ &  $6.56 \pm 3.43 \pm 3.31$ &  $234.9 \pm 19.5\pm 13.3$
  & $0.136\pm 0.030 \pm 0.035$  \\
  \hline \hlinestrut
    Fit C.L & 3.0\% & & \\
  \hline
  \end{tabular}
\end{table}

%
%
\subsubsection{\boldmath Results for the $D^+\to\pi^+\pi^-\pi^+$ decay}

\begin{figure}[htbp]
  \centering
  \subfigure
  {%
    \epsfig{file=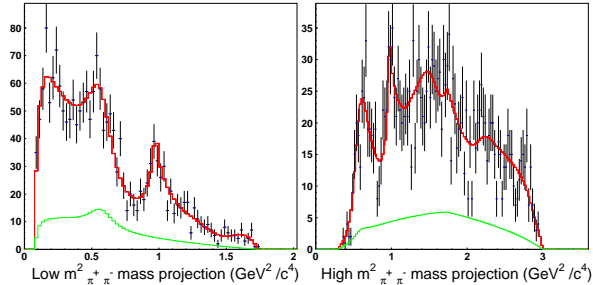,width=0.5\linewidth}
  }
  \caption{$D^+$ Dalitz-plot projections with the final fit
  superimposed. The background shape under the signal is also shown.}
  \label{dp_proj_Kmatrix}
\end{figure}

The $D^+\to\pi^+\pi^-\pi^+$ Dalitz plot shows an excess of events at low
$\pi^+\pi^-$ mass, which cannot be  explained in the context of the simple
isobar model with the usual mixture of well established resonances along with a
constant, non-resonant amplitude. A new scalar resonance, the $\sigma(600)$,
has been previously proposed \cite{e791_dp} to describe this excess. However we
know that complex structure can be generated by the interplay among the
$S$-wave resonances and the underlying non-resonant $S$-wave component that
cannot be properly described in the context of a simple isobar model. It is
therefore interesting to study this channel with the present formalism, which
embeds all the experimental knowledge about the $S$-wave $\pi^+\pi^-$
scattering dynamics. With the same procedure based on statistical significance
and fit confidence level used in the $D_s^+$ analysis, the final set of
contributions is reached. Beside the $S$-wave component, the decay appears to
be dominated by the $\rho^0(770)$ plus a $f_2(1270)$ component. The
$\rho^0(1450)$ was always found to have less than $2\,\sigma$ significance and
was therefore dropped from the final fit. In analogy with the $D_s^+$, the
direct three-body non-resonant component was not necessary since the SVP of the
$S$-wave could reproduce the entire non-resonant portion of the Dalitz plot.
The complete fit results are reported in Table~\ref{table_dp_kmatrix}.
\begin{table}[htb]
  \centering
  \caption{Fit results from the \emph{K-matrix} model fit for $D^+$.}
  \label{table_dp_kmatrix}
  \begin{tabular}{cccc}
  \hline \hlinestrut
  decay channel  & fit fraction (\%) & phase (deg) & amplitude coefficient \\
  \hline \hlinestrut
    ($S$-wave)\,$\pi^+$   &  $56.00 \pm 3.24 \pm 2.08$  & 0
  (fixed)
  &  1 (fixed)    \\
  $f_2(1270)\,\pi^+$   & $11.74 \pm 1.90 \pm 0.23$ & $-47.5 \pm 18.7\pm 11.7$ &
  $1.147\pm 0.291 \pm0.047$ \\
  $\rho^0(770)\,\pi^+$ & $30.82 \pm 3.14 \pm 2.29$ &$-139.4 \pm 16.5 \pm 9.9$ &
  $1.858 \pm 0.505 \pm0.033$
    \\
  \hline \hlinestrut
   Fit C.L. & 7.7\% & & \\
  \hline
  \end{tabular}
\end{table}
The $D^+$ Dalitz projections are shown in Fig.~\ref{dp_proj_Kmatrix}. The fit
C.L. is 7.7\%.
%
%

The most interesting feature of these results is the fact that the better
treatment of the $S$-wave contribution provided by the \emph{K-matrix} model
can reproduce the low-mass $\pi^+\pi^-$ structure of the $D^+$ Dalitz plot.
This suggests that any $\sigma$-like object in the $D$ decay should be
consistent with the same $\sigma$-like object measured in the $\pi^+\pi^-$
scattering. Additional studies with higher statistics will be required to
completely understand the $\sigma$ puzzle. It is interesting to recall the
close analogy between the $D\to\pi\pi\pi$ channel and the $B\to\rho\pi$ one,
which is a good candidate to measure the angle $\alpha$ of the Standard Model
Unitarity Triangle; the analysis of
$B\to\rho\pi$ will proceed through a time-dependent Dalitz-plot analysis of the
three-pion final state and will likely present similar parametrization
complications.

\subsection{ \boldmath Interpretation of the $D^+_s$ and $D^+\to\pi^+\pi^-\pi^+$ results}

The \emph{K-matrix} formalism has been applied for the first time to the charm
sector in the FOCUS Dalitz-plot analyses of the $D_s^+$ and
$D^+\to\pi^+\pi^-\pi^+$ final states. The results are extremely encouraging since
the same \emph{K-matrix} description gives a coherent picture of both two-body
scattering measurements in light-quark experiments \emph{as well as} charm meson
decay. This result was not obvious beforehand. Furthermore, the same model is able
to reproduce features of the $D^+\to\pi^+\pi^-\pi^+$ Dalitz plot that otherwise
would require an \emph{ad hoc} $\sigma$ resonance. In addition, the non-resonant
component of each decay seems to be described by known two-body $S$-wave dynamics
without the need to include constant amplitude contributions. The \emph{K-matrix}
treatment of the $S$-wave component of the decay amplitude allows for a direct
interpretation of the decay mechanism in terms of the five virtual channels
considered:  $\pi\pi$, $K\bar K$, $\eta\eta$, $\eta\eta '$ and $4\,\pi$. By
inserting $KK^{-1}$ in the decay amplitude, $F$,
\begin{equation}
  F = (I-iK\rho)^{-1}P=(I-iK\rho)^{-1}KK^{-1}P=TK^{-1}P=TQ
  \label{q-vect}
\end{equation}
we can view the decay as consisting of an initial production of the five
virtual states which then scatter via the physical $T$ into the final state.
The \emph{Q-vector} contains the production amplitude of each virtual channel
in the decay. The resulting picture, for both $D_s^+$ and $D^+$ decay, is that
the $S$-wave decay is dominated by an initial production of $\eta\eta$,
$\eta\eta'$ and $K\bar K$ states. Dipion production is always much smaller.
This suggests that in both cases the $S$-wave decay amplitude primarily arises
from a $s\bar s$ contribution such as that produced by the Cabibbo-favored weak
diagram for the $D_s^+$ and one of the two possible singly Cabibbo-suppressed
diagrams for the $D^+$. For the $D^+$, the $s\bar s$ contribution competes with
a $d\bar d$ contribution. That the $f_0(980)$ appears as a peak in the $\pi\pi$
mass distribution in $D^+$ decay, as it does in $D_s$ decay, shows that for the
$S$-wave component the $s{\bar s}$ contribution dominates~\cite{penn1}.
Comparing the relative $S$-wave fit fractions that we observe for $D_s^+$ and
$D^+$ reinforces this picture. The  $S$-wave decay fraction for the $D_s^+$
(87\%) is larger than that for the $D^+$ (56\%). Rather than coupling to an
$S$-wave dipion, the $d\bar d$ piece prefers to couple to a vector state like
$\rho^0(770)$ that alone accounts for $\sim 30$ \% of $D^+$ decay. This
interpretation also bears on the role of the annihilation diagram in the
$D_s^+\to\pi^+\pi^-\pi^+$ decay. This study suggests that the $S$-wave
annihilation contribution is negligible over much of the dipion mass spectrum.
It might be interesting to search for annihilation contributions in higher spin
channels, such as $\rho^0(1450)\pi$ and $f_2(1270)\pi$.

\section{Conclusions}

Dedicated studies of charm physics are foreseen in the near future at existing
collider facilities such as CLEO-c, the $B$-factories, and the Tevatron. Other
facilities might also play a role, such as GSI and Super $B$-factories. In
parallel, the mature stage of the ongoing analysis is providing warnings and
lessons on how to perform proper studies to interpret and understand Heavy
Flavor phenomenology and, hopefully, to find signs of New Physics.

\end{document}